\documentclass{article}

\usepackage[utf8]{inputenc}
\usepackage[T1]{fontenc}
\usepackage{amssymb,amsmath,amsthm,amsfonts}
\usepackage[caption=false,font=normalsize,labelfont=sf,textfont=sf]{subfig}

\usepackage{todonotes}
\usepackage{cleveref}
\usepackage{soul}
\usepackage{paralist}
\usepackage{xspace}
\usepackage{multirow}
\usepackage{fullpage}
\usepackage{algorithm}
\usepackage{float}
\usepackage[noend]{algpseudocode}
\usepackage{authblk}
\usepackage{xurl}
\usepackage{eurosym}
\newcommand{\mone}{FCSE\xspace}
\newcommand{\mtwo}{BCSE\xspace}
\newcommand{\Methodone}{Forward checking soiling estimator\xspace}
\newcommand{\Methodtwo}{Backward checking soiling
estimator\xspace}

\renewcommand\st[1]{}
\renewcommand\todo[1]{}

\DeclareMathOperator*{\mede}{mede}

\begin{document}

\title{Data-driven soiling detection in PV modules}

\author[1]{Alexandros Kalimeris}

\author[1]{Ioannis Psarros}

\author[1]{Giorgos Giannopoulos}

\author[1]{Manolis Terrovitis}

\author[1]{George Papastefanatos}

\author[2]{Gregory Kotsis}
\affil[1]{Athena RC}
\affil[2]{INACCESS Networks}

\maketitle

\begin{abstract}
Soiling is the accumulation of dirt in solar panels which leads to a decreasing trend in solar energy yield and may be the cause of vast  revenue losses. 
The effect of soiling can be reduced by washing the panels, which is, however, a procedure of non-negligible cost. Moreover, soiling monitoring systems are often unreliable or very costly. 
We study the problem of estimating the soiling ratio in photo-voltaic (PV) modules, i.e., the 
ratio of the real  power output to the  power output
that would be produced if solar panels were clean. A key advantage of our algorithms is that they estimate soiling, without needing to train on labelled data, i.e., periods of explicitly monitoring the soiling in each park, and without relying on generic analytical formulas which do not take into account the peculiarities of each installation. 
 We consider as input a time series 
comprising a minimum set of measurements, that  are available to most PV park operators. 
Our experimental evaluation shows that we significantly outperform current state-of-the-art methods for estimating soiling ratio.


\end{abstract}

\section{Introduction}
Soiling is the accumulation of dirt on the surfaces of photo-voltaic (PV) modules, which leads to a loss in the power output. Soiling is typically caused by airborne particles, including for example dust, pollen and soot. Depending on the location, soiling may also be caused by heavier material such as ice, bird droppings, or falling leaves. 

One standard way to quantify soiling is by the \emph{soiling ratio} $SR$~{{\cite{iec}}}, which is defined as the 
ratio of the real  power output to the power output
that would be produced if solar panels were clean. \emph{Soiling loss} is then  defined as $1-SR$, and  \emph{soiling
rate} is defined as the (daily)  rate of change of the soiling loss. Other metrics have been also proposed, e.g., the insolation-weighted soiling
ratio~\cite{DMM18}, aiming to better capture the loss induced by soiling. 

To reduce the effect of soiling, PV modules must be cleaned on strategically chosen dates to reduce the cost induced by energy loss while taking into account cleaning costs. Detection of time periods during which soiling severely affects power output is therefore significant for the efficient scheduling of cleanings. What makes the problem challenging is the shortage of labelled data which is caused by the fact that soiling monitoring systems are often considered unreliable or costly. {For example, soiling stations which are the most common commercially available soiling monitoring solution~{\cite{BESSA2021102165}}, still require regular cleanings and maintenance, which can be expensive, especially in remote locations, and imperfect cleanings can result in 
significant measurement uncertainty~{\cite{8366214}}.}  Therefore, soiling periods must be deduced from measurements of a  number of reliable variables, e.g., power output, irradiance, temperature.

Existing methods that detect soiling follow two alternative strategies: a) they train a model on labelled data, i.e., data where the soiling of the panels has been logged using specialized sensors and cleaning events have been explicitly recorded (e.g.,~\cite{MASSIPAVAN20111128,MPMDPK13}) and b) by using an analytical formula for optimal energy output based on environmental readings (e.g.,~\cite{4060159,DMM18, MTLSGGMAF21}). The former strategy is more accurate but requires significant resources to produce the labelled data, which must be produced for each different installation. The latter strategy does not take into account the peculiarities of each installation and leads to less accurate results (as we demonstrate in \Cref{section:experiments}). \st{The main advantage of our method, being purely data driven, is that it is able to effectively identify periods of optimal performance, without the use of labelled data, and utilize these periods as reference for detecting soiling.} 
{The main advantage of our method, is that it is purely data-driven, in the sense that it does not require a generic analytical formula for the relation between power output and the commonly used environmental readings, but it learns this relation in a self-supervised manner (without the need for labelled data).}
This way we achieve better results than methods that rely on analytical formulas without the cost of methods that need explicitly labelled data. 

We consider as input the monitoring data from the park operation, i.e., a time series with measurements of power output, irradiance, and  module temperature for a certain array or string of PV modules,  precipitation, and dates on which the solar panels were manually cleaned for maintenance (if such information exists). 
The soiling ratio over a sequence of timestamps $t_1,\ldots,t_n$ is defined as 
$\mathcal{SR} = \frac{P_{t_1}}{{P}_{t_1}^{\ast}},\ldots, \frac{P_{t_n}}{{P}_{t_n}^{\ast}}$,  where each $P_{t_i}$ is the actual power output corresponding to timestamp $t_i$, and ${P}_{t_i}^{\ast}$ is the expected power output assuming that the solar panels are clean, corresponding to the same timestamp. 
Our framework trains a regression model $\mathcal{M}$ which accurately predicts ${P}_{t_1}^{\ast},\ldots,{P}_{t_n}^{\ast}$ (which are not given as input). This yields an estimate for the soiling ratio as  $\mathcal{SR}_{\mathcal{M}}=\frac{P_{t_1}}{\tilde{P}_{t_1}},\ldots, \frac{P_{t_n}}{\tilde{P}_{t_n}}$,  where each 
$\tilde{P}_{t_i}$ is the value predicted by  $\mathcal{M}$ for timestamp $t_i$. We aim for $\mathcal{M}$ such that $\mathcal{SR}_{\mathcal{M}} \approx \mathcal{SR}$. 
Raining periods (extracted from precipitation measurements), and  manual cleanings, are used in the ``learning'' phase of our proposed model. One of our methods can run exclusively on rain information, in case manual cleanings are not performed or logged. 
Our approach is robust to misinformation about manual cleanings because it checks each potential cleaning to determine its effect on power output. Manual cleanings that are not logged, have a negligible effect; 
they can only affect the quality of the training set positively. 

The main advantages of our method are that they do not require measurements of soiling from specialized equipment which can be costly or inaccurate, they do not rely on the accuracy of an analytical formula for the optimal energy output of the park, and they agnostic to the type of PV modules employed. 
As a purely data-driven approach, it solely depends on the availability of data, and in particular a minimal set of generally available variables. 
Our approach is robust to misinformation about manual cleanings because it checks each potential cleaning to determine its effect on power output. Moreover, manual cleanings that are not logged, have a negligible effect on our approach; 
their existence can only affect the quality of the training set positively. 


In \Cref{section:related}, we discuss related work, in \Cref{section:preliminaries} we provide necessary background,  in \Cref{section:soiling} we present a detailed description of our methods, and 
in \Cref{section:experiments} we present  our experimental findings.

\section{Related work}
\label{section:related}



PVUSA introduced a method for rating PV systems based on a simple regression model \cite{PVUSA} which employs the simplified assumption that  array current depends only on
irradiance and that array voltage depends only on module temperature.  
Massi Pavan et al.~\cite{MASSIPAVAN20111128}  compare the standard test conditions (STC) (irradiance: $1000 W/m^2$, module temperature: $25^{\circ} C$) performance of a PV park before and after its cleaning.
In order to determine the performance at STC conditions they use a regression model, suggested in \cite{MWPP08},  that accepts as input the two main  climate features, i.e. the in-plane global irradiance and the photo voltaic module temperature. 
However, their work requires as input labelled data, i.e. time series extracted from both clean and soiled PV modules. 
Massi Pavan et al.~\cite{MPMDPK13} developed four Bayesian Neural Network (BNN) models with the aim to calculate the STC performance of two plants before and after a complete clean-up of their modules. The idea is that differences between the STC power before and after the clean-up represent the losses due to the soiling effect. However, their work also requires as input labelled data, i.e. time series extracted from both clean and soiled PV modules.


Closer to our work are methods which estimate soiling losses based on PV system data.
The  Fixed Rate Precipitation (FRP) method~\cite{4060159} calculates the daily soiling loss. The method requires as input:  the slope of the performance metric/index during the longest dry period, a cleaning threshold for rains, i.e.,  the minimum amount of daily precipitation required to have a cleaning effect on PV modules, and a number of days after a raining period for which no soiling occurs. The method implicitly assumes that the soiling rate  remains the same throughout time.  This requirement can be very restrictive, because of the different types of soiling that may occur, depending also on the location or the season. For the same reason, it is unrealistic to assume that there is a certain minimum value classifying rains as effective. More recently, 
 Deceglie, Micheli, and Muller~\cite{DMM18} developed a new method for quantifying soiling loss, which compares favourably to FRP. The new method is termed 
the stochastic rate and recovery
(SRR) method. 
It uses an analytical formula, calculated over values for irradiance and module temperature, to compute the expected power output, which is then used to compute a performance metric.  
The method  
first detects soiling intervals in a dataset, and then, based on the observed characteristics of each interval, estimates the total loss. Notice that SRR provides an aggregate estimate of soiling loss, calculated for the whole input period, while our focus lies on determining soiling loss even on shorter periods of time. 
Skomedal and  Deceglie~\cite{SD20} proposed the combined degradation and soiling  method for further analyzing a performance metric signal. 
Finally, Micheli et al.~\cite{MTLSGGMAF21} consider non-linear degradation in soiling intervals, and they apply various methods for changepoints detection to obtain a refined soiling profile. 
All methods studied there are based on finding changepoints on the performance metric curve, as calculated by SRR. On the contrary, our approach detects changepoints as an intermediate step towards computing a performance metric.  It is apparent from recent work that improvements on estimating the expected power output directly translate to improvements on various tasks in PV data analysis.  

\section{{Methodology}}
\subsection{Preliminaries}
\label{section:preliminaries}
\subsubsection{Basic assumptions and definitions}
\label{subsection:definitions}

 Our input consists of a multi-variate time series containing measurements for: \begin{inparaenum}[i)]
    \item power output,
    \item irradiance,  
    \item module temperature, 
    \item precipitation.
\end{inparaenum}
Our methods can be further enhanced if we are also given as input the dates on which the PV modules were manually cleaned. 

Let $\mathcal{R}$ be the set  of all rains, defined as follows: $[t,t']\in \mathcal{R}$ if and only if there is a rain starting at $t$ and ending at $t'$. Rains are extracted from input as maximal time intervals containing  positive precipitation values. Similarly, if manual cleanings are provided let $\mathcal{C}$ be the set of all such intervals, defined as follows: $[t,t']\in \mathcal{C}$ if and only if we know that the PV modules were being cleaned between timestamps $t$ and  $t'$.
 We denote by $\mathcal{W}_p$ the set 
 of all potential cleaning events, defined as $\mathcal{W}_p = \mathcal{C} \cup \mathcal{R}$. 
We assume that precipitation measurements are sufficiently frequent, 
so that we can accurately detect rains. 

\subsubsection{Regression models}
\label{subsection:regression}


A basic component of our methods is regression. We fit regression models to  represent power output during ``dirty'' or  ``clean'' periods and we use prediction errors to detect performance changes. We consider as feature variables the irradiance and the module temperature, and the target outcome corresponds to the power output. We apply  \textit{Ridge Regression with polynomial features}, which is parameterized by the  degree of the regression polynomial, and
a regularization strength parameter for the linear least squares function 
 (the loss function) where regularization is given by the $\ell_2$-norm. 
  The parameters were selected during the initial stages of the algorithm development process, where we experimented with cross-validation and hyper-parameter tuning techniques. The exact values used in our experiments are discussed in \Cref{section:experiments}. 
Our model selection was a consequence of preliminary experiments with various {(simple)} regression models such as Ordinary Least Squares, Support Vector Regression, etc.{, that we executed in a CPU with maximum processor frequency at 3.7GHz, and available RAM at 256Gb. In the experiment that we conducted, we randomly choose $100$ time intervals of maximum duration of one month from the time series provided in~{\cite{dataset}}, which are also discussed in Section~{\ref{section:experiments}}, and we randomly split them into  training and testing subsets containing $80\%$ and $20\%$ of the points respectively.  
Our choice satisfies a bifold objective: i) good accuracy and ii) fast fitting time. The latter is vital in our method which fits one model for each potential cleaning. Table~{\ref{table:modeleval}} contains MAPE values and fitting times for four different models. Polynomial features and the polynomial kernel used in Support Vector Regression (SVR) are of degree $3$. The highest accuracy is achieved by SVR with linear kernel and polynomial features, being roughly $0,4\%$ better than Ridge Regression which is the second best. However, the fitting time of SVR is at least one order of magnitude higher than that of Ridge Regression. } 
Ridge Regression 
is a \st{relatively} simple model that adds only one extra tunable parameter to our learning pipeline, and the regularization it provides acts as a measure to prevent overfitting. We \st{should} also emphasize the fact that one can easily plug-in any regression model in our approach. 



\begin{table}[h]
\centering
 \caption{Evaluation of regression models.} 
 \label{table:modeleval}
\begin{tabular}{|l|l|l|}
\hline
\textbf{Model}                                                                                                  & \textbf{MAPE} & \textbf{Fitting time (s)} \\ \hline
\begin{tabular}[c]{@{}l@{}}Linear Regression \\ with polynomial features\end{tabular}                           & 0.0812        & 0.0015                    \\ \hline
\begin{tabular}[c]{@{}l@{}}Ridge Regression \\ with polynomial features\end{tabular}                            & 0.0807        & 0.0012                    \\ \hline
\begin{tabular}[c]{@{}l@{}}Support Vector Regression \\ with polynomial kernel\end{tabular}                     & 1.0648        & 0.0177                    \\ \hline
\begin{tabular}[c]{@{}l@{}}Support Vector Regression \\ with linear kernel and polynomial features\end{tabular} & 0.0770        & 0.0666                    \\ \hline
\end{tabular}
\end{table}


Several steps in our approach rely on computing  measures for the prediction accuracy of our model. Let $\mathcal{Y}=Y_{t_1},\ldots, Y_{t_n}$,  $\tilde{\mathcal{Y}}=\tilde{Y}_{t_1'},\ldots, \tilde{Y}_{t_n'}$ be two univariate time series, and let $T=\{t_1,\ldots,t_n\},T'=\{t_1',\ldots,t_n'\} $.  
We use a variant of the mean absolute percentage error (MAPE) which is defined over time intervals as follows: for any $[t,t'] \subseteq T\cap T' $,
\[\mathrm{mape}_0(\mathcal{Y}, \tilde{\mathcal{Y}}, [t,t']) =
\frac{\mathrm{mean}(\{|Y_j-\tilde{Y}_j| \mid j\in  [t,t']\}}{\mathrm{mean}(\{|Y_j| \mid j\in  [t,t']\})}.\]
Note that $\mathrm{mape}_0$ is robust to zero true values (as long as not all of them are zeroes) since it uses as denominator the mean of the values, as opposed to standard MAPE where all actual values appear as  denominators leading to singularities even if there is only one zero true value.  When $\mathcal{Y}$ and $\tilde{\mathcal{Y}}$ are clear from the context, we omit them from our notation and we simply write $\mathrm{mape}_0( [t,t'])$. We also use the median multiplicative error defined as
$
\mathrm{mede}(\mathcal{Y}, \tilde{\mathcal{Y}}) =
\mathrm{median}\left(\left\{\frac{Y_{i}}{\tilde{Y}_{j}}\mid i\in T, j\in T'\right\}\right) $.  

\subsection{Soiling detection}
\label{section:soiling}
\begin{figure*}[h]
     \centering
     \includegraphics[width=\textwidth]{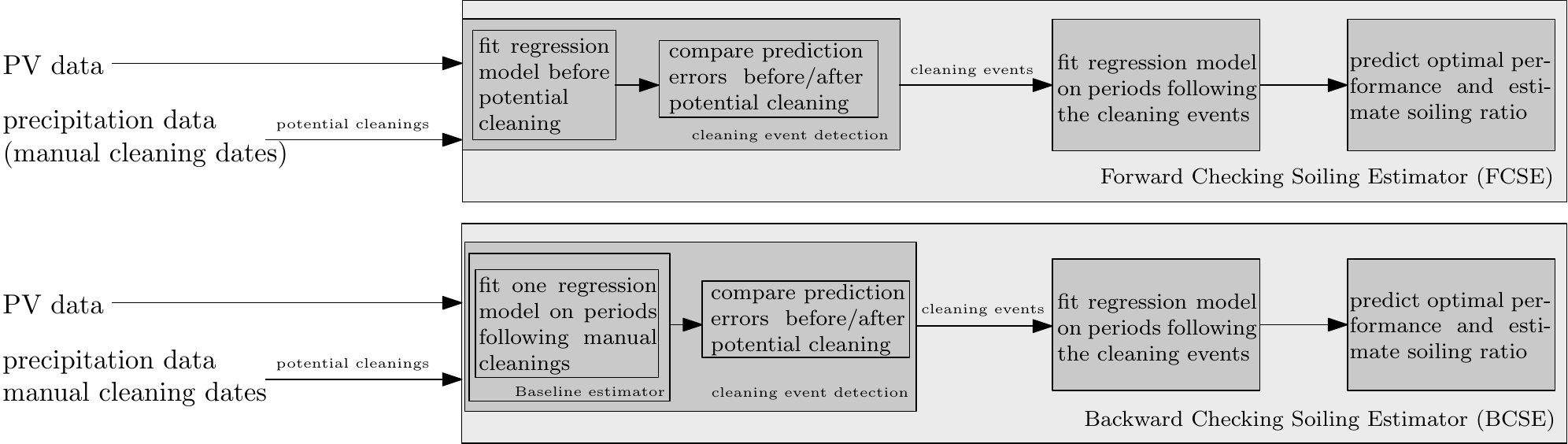}
     \caption{Basic steps of our methods. Manual cleanings are optional for FCSE. To detect cleaning events, FCSE fits one regression model before each potential cleaning event, while BCSE fits one regression model using manual cleaning dates and uses it in classifying all cleaning events. }
     \label{fcse_bcse}
 \end{figure*}

In this section, we formally describe our methods, which are composed of two main steps. The first step is that of  detecting cleaning events. Then, using these cleaning events we define training periods for regression models aiming to capture the optimal performance of the PV modules. 
\st{Measurements of power output, irradiance, and module temperature are scaled to $[0,1]$ by subtracting the minimum value and dividing by the range of values. }
In all our methods, we fit regression models which capture the dependence of power output on the values of irradiance and module temperature, i.e., power output is the dependent variable, while irradiance and module temperature are the feature variables. {Measurements are scaled to $[0,1]$ by subtracting the minimum value and dividing by the range of values.} {Figure~{\ref{fcse_bcse}} summarizes the main steps of our methods.}


 
 

\subsubsection{Baseline soiling estimator}
 We first present our baseline approach for estimating the soiling ratio. Our baseline algorithm is based on the following assumption: manual cleanings alone  define points in time where the PV modules are clean. While these points are not sufficiently many to define a training set, we can extend them to short intervals of a user-defined length $w_{train}$. 
This is the amount of time during which we can safely assume that the panels remain clean. 

We fit a regression model that aims to capture the power output when PV modules are clean. To this purpose, we fit a regression model $\mathcal{M}$ on the set  of input points with timestamps from $\bigcup_{[t,t']\in \mathcal{C}}[t',t'+w_{train}]$. 
We define 
$\mathcal{SR}_{\mathcal{M}}=\frac{P_{t_1}}{\tilde{P}_{t_1}},\ldots, \frac{P_{t_n}}{\tilde{P}_{t_n}}$ as the modelled soiling ratio where each $P_{t_i}$ is an input power output value, and $\tilde{P}_{t_i}$ is the value predicted  $\mathcal{M}$.

\subsubsection{\Methodone (\mone)}
 Our first method examines each potential cleaning event independently and assigns scores which represent the significance of the detected change of behavior. Five input parameters are required: the length of the training period  $w_{1}$, the length of the validation period $w_{2}$, the length of the test period $w_{3}$, a parameter $q$ defining the  quantile of the scores which classifies events as cleanings, and the length $w_{train}$ defining the training set for the final regression model used to estimate soiling. 
For each interval $[t,t']\in \mathcal{W}_p$
, we   
 fit a regression model in the time interval $[t-w_1-w_2,t-w_2)$,  we validate it in the time interval $[t-w_2,t)$ and we test it in the time interval $(t',t'+w_3]$. We compute the function $\mathrm{mape}_0$ on the validation interval and if the returned value is greater than $5\%$ then we consider this event invalid and we discard it from further consideration. This threshold aims to discard events that we are unable to classify with  certainty. 
\st{Our choice of $5\%$ is the result of experimenting with real data, but  tighter thresholds may be more appropriate in other settings.} {The reasons behind choosing $5\%$ as our threshold are the following. First, due to the nature of our task, the regression model is required to make very accurate predictions and detect power deviations at a very small scale. This requires high accuracy of our regression models; therefore a tight threshold. On the other hand, this threshold must be pragmatic: having an extremely small value as a threshold will lead to unrealistic outputs where no cleaning events are detected and, consequently, no soiling estimation can be derived. }{ We experimentally validate our choice of $5\%$ in Section~{\ref{ss:mapethreshold}}}.  

The intuition is that if the PV modules under-perform due to soiling, for a time period preceding $t$, then the regression model captures this under-performing behaviour and if $[t,t']$ is a cleaning event then the model should underestimate the power output in $(t',t'+w_3]$. To compute the score 
 of the potential cleaning event $[t,t']$, we first compute $\mathcal{PI}_{val}$ as the sequence of actual power output values divided by the predicted power output values for the time interval $[t-w_2,t)$, and $\mathcal{PI}_{test}$ as the sequence of actual power output values divided by the predicted power output values for the time interval $(t',t'+w_3]$. Then, the score assigned to $[t,t']$ is $\mede(\mathcal{PI}_{val}, \mathcal{PI}_{test})$. 
 We define as cleaning events all intervals $[t,t']\in \mathcal{W}_p$ with score above the $q$th-quantile of all scores. Let $\mathcal{W}_1$  be  the set of detected cleaning events. We fit a regression model $\mathcal{M}$ on the input points with timestamps from $\bigcup_{[t,t']\in \mathcal{W}_1}[t',t'+w_{train}]$. The intuition is that cleaning events define points in time where the PV modules are clean. Obviously, these points are not sufficiently many to define a proper training set. By extending these points to (short) intervals, of length $w_{train}$, we increase the size of the training set without (significantly) affecting its quality.  
We define $\mathcal{SR}_{\mathcal{M}}=\frac{P_{t_1}}{\tilde{P}_{t_1}},\ldots, \frac{P_{t_n}}{\tilde{P}_{t_n}}$ as the estimated soiling ratio where each $P_{t_i}$ is an input power output value, and $\tilde{P}_{t_i}$ is the value predicted by the regression model $\mathcal{M}$.

 Notice that \mone does not require having the cleaning dates $\mathcal{C}$ as input, and we could simply have $\mathcal{W}_p=\mathcal{R}$. 
 
 \subsubsection{\Methodtwo (\mtwo)}

 Our second method builds upon the baseline approach. This method requires five input parameters $w_1$, $w_2$, $w_3$, $q$, $w_{train}$. Parameters $w_1$ and $w_2$ denote the length of the testing period preceding the potential cleaning event and the length of the validation period following the potential cleaning event respectively. 
 Parameter $w_3$ denotes the length of the time period following each $[t,t']\in \mathcal{C}$ such that the modules remain clean. 
Parameter $q$ defines the  quantile of the scores which classifies events as cleanings. 
Parameter $w_{train}$ is used to define the training set of the final regression model for estimating the soiling ratio. 
 We train one regression model on the set of points defined by timestamps in  $\bigcup_{[t,t']\in\mathcal{C}} [t',t'+w_3]$. This model aims to capture modules' ``clean'' performance. For each $[t,t']\in \mathcal{W}_p$, we use our model to make predictions on $[t-w_1,t)$ and $(t',t'+w_2]$. If  $\mathrm{mape}_0((t',t'+w_2])$ is greater than $5\%$ then we consider this interval invalid and we discard if from further consideration. As in \mone, this filtering step is to avoid considering events that our models fail to classify with a good amount of certainty. 
 
  The intuition is that if $[t,t']$ is a cleaning event, then the PV modules' performance during $[t',t'+w_2]$ must resemble the ``clean'' performance as predicted by our regression model. Similarly, if the modules under-perform during $[t-w_1,t)$, then the induced ratio of the actual power output over the predicted power output must be significantly smaller than $1$. To compute the score 
 of the potential cleaning event $[t,t']$, we first compute $\mathcal{PI}_{before}$ as the sequence of actual power output values divided by the predicted power output values for the time interval $[t-w_1,t)$, and $\mathcal{PI}_{after}$ as the sequence of actual power output values divided by the predicted power output values for the time interval $(t',t'+w_2]$. Then, the score assigned to $[t,t']$ is $\mede(\mathcal{PI}_{before}, \mathcal{PI}_{after})$. 
  We define as our threshold parameter $thrsh$ the $q$th-quantile of all scores. 
We define as cleaning events  all intervals $[t,t']\in \mathcal{W}_p$
with score above the $q$th-quantile of all scores. Let $\mathcal{W}_2$,  be  the set of detected cleaning events. We fit a regression model $\mathcal{M}$ on the input points with timestamps from $\bigcup_{[t,t']\in \mathcal{W}_2}[t',t'+w_{train}]$. As in \mone, the intuition is that cleaning events define points in time where the PV modules are clean. Obviously, these points are not sufficiently many to define a training set. By extending these points to (short) intervals, of length $w_{train}$, we increase the size of the training set without (significantly) affecting its quality.  
We define $\mathcal{SR}_{\mathcal{M}}=\frac{P_{t_1}}{\tilde{P}_{t_1}},\ldots, \frac{P_{t_n}}{\tilde{P}_{t_n}}$ as the estimated soiling ratio where each $P_{t_i}$ is an input power output value, and $\tilde{P}_{t_i}$ is the value predicted by  $\mathcal{M}$.


 \section{Experiments}
 \label{section:experiments}

 \subsection{Datasets}
 
 
\paragraph{State-of-the-art dataset}
\label{sss:evaldata}
 To evaluate our methods, we use a dataset provided in
 \cite{dataset}, which contains a set of current-voltage (I-V) curves and associated meteorological data for PV modules representing all flat-plate PV
technologies and for three different locations and climates for approximately one-year periods. 
\st{For each location, we are given values for a normalized metric, called \emph{soiling derate}, which compares the daily performance of a
PV module to an identical PV module that is cleaned
during daily maintenance. This is computed using measurements for short-circuit current and irradiance  from these two
identical PV modules (cleaned and not cleaned). 
Soiling derate is the result of dividing daily values of ampere-hours per kilowatt-hours per square
meter Plane of Array (POA) irradiance for the not-cleaned PV module, by the corresponding values of the cleaned PV module.} 
{For each location, we are given values for a normalized metric, called \emph{soiling derate} which is computed using measurements for short-circuit current and irradiance  from two identical PV modules; one that is cleaned during daily maintenance, and one that is not. Soiling derate is the result of dividing daily values of ampere-hours per kilowatt-hours per square
meter Plane of Array (POA) irradiance for the not-cleaned PV module, by the corresponding values of the cleaned PV module~{\cite{dataset}}. 
The soiling derate aims to provide a performance index analogous to soiling ratio, estimated on real measurements.} We emphasize that soiling derate is only used for the evaluation of our methods 
and are not utilized as input (nor in SRR). 
The time granularity is $5$ minutes, and measurements are provided for all hours of daylight.  The three locations are Cocoa, Florida, USA; Eugene, Oregon, USA; and Golden, Colorado, USA. PV modules in Cocoa and Eugene were cleaned when this was necessary in order to ensure that levels of soiling loss were maintained at a reasonable level. PV modules in Golden were not cleaned because frequent rains helped maintaining a reasonable level of soiling loss. Cocoa has a minimum soiling derate of 0.985, 
Eugene has a minimum soiling derate of 0.964, and Golden has a minimum soiling derate of 0.977.

 In our methods, we use measurements for the maximum power of the PV module in watts, the amount of solar irradiance in watts per square meter
received on the PV module surface, the PV module back-surface temperature and the accumulated daily total precipitation. The dataset also provides dates on which  all PV modules were cleaned. We apply our methods on  PV modules that were used in estimating the soiling derate, and in particular on those that were not cleaned every day.
 As discussed in \Cref{subsection:regression} our methods utilize Ridge Regression models. For those models, we use polynomial features of the $3$rd degree and a regularization strength parameter $alpha = 10^{-4}$ during the fitting stages.
 
  \paragraph{Real-world dataset}
 \label{sss:inaccessdata}
 We also consider a real-world 
scenario, where no ground truth is available. We test our methods
on  a  dataset from a very different location and of different climate conditions, comprising measurements from a solar park located in Greece. We are given values for power output, irradiance, module temperature and precipitation on a time granularity of 15 min for a period of approximately 7 years, and 15 dates of manual cleanings. 
 

 \subsection{Method evaluation and discussion}
\subsubsection{Soiling estimation}
\label{sss:soiling}
 We evaluate our methods, by comparing them to the analogous  model used in SRR. {To show robustness of our methods in different parameter settings, we try various  lengths for the periods used in changepoint detection. Table~{\ref{table:rmse1}} lists the respecting values (in days) for parameters $w_1,w_2,w_3$ in {\mone} and $w_1,w_2$ in {\mtwo}. The rest of the parameters are set as follows: we  apply {\mone} with parameters $q=0.9$, and $w_{train}=30$ days and  {\mtwo} with parameters $q=0.9$, $w_3=30$ days,  and $w_{train}=30$ days.  The baseline soiling estimator is applied with $w_{train}=30$ days. Since our methods are unsupervised, classic automated methods fail to optimize the above parameters. Essentially, domain expertise is the main lead for selecting  parameters appropriately,  also depending on the properties of each location that affect the rate at which soiling progresses. 
 However, as Table~{\ref{table:rmse1}} indicates, the methods are robust within a range of reasonable values for the parameters. The fixed parameters $w_{train}$ (and $w_3$ in {\mtwo}) define time periods during which a clean solar panel is likely to remain clean. While smaller values for $w_{train}$ (resp.~$w_3$) seem to provide safer conclusions, larger values provide a bigger size and diversity of the induced training set. The parameter $q$ defines a threshold on how important a changepoint should be to be considered as a cleaning event. Setting $q=0.9$ implies that the top-scored $10\%$ of potential cleanings will be considered as cleaning events. Factors that must be taken into account when setting this parameter include the total number of potential changepoints, parameters $w_3$, $w_{train}$, and the size of the dataset. While larger values of $q$ tend to lead to safer conclusions about cleaning events, this may lead to a decreased size of the training set, negatively affecting the final regression model. } 
\st{We  apply {\mone} with parameters $w_1=10$ days, $w_2=5$ days, $w_3=10$ days, $q=0.9$, and $w_{train}=30$ days and  {\mtwo} with parameters  $w_1=5$ days, $w_2=10$ days $w_3=30$ days,  and $w_{train}=30$ days.  The baseline soiling estimator is applied with $w_{train}=30$ days.
}

 \begin{figure*}[h]
     \centering
     \includegraphics[width=\textwidth]{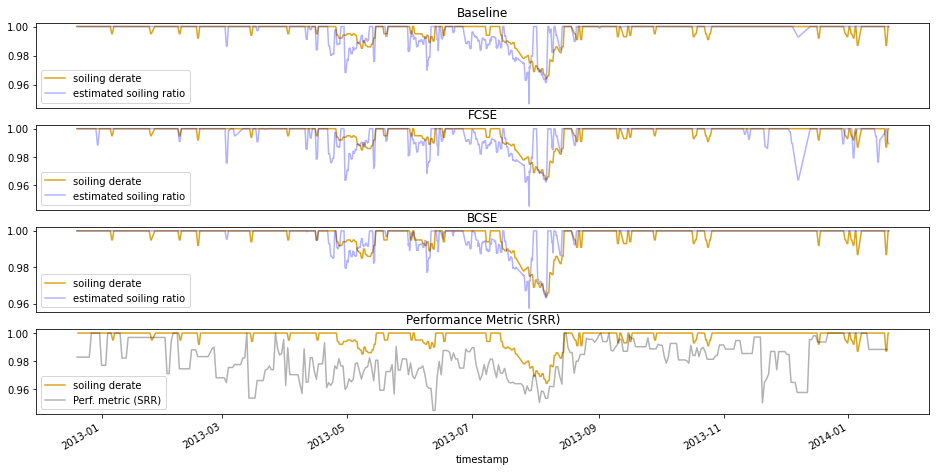}
     \caption{Soiling ratio predicted by our models, and the performance metric used in SRR, for the Eugene dataset. {{\mone} with parameters $w_1=10,w_2=5,w_3=10$ and {\mtwo} with parameters $w_1= 5, w_2=10$.}}
     \label{fig:soilingratio}
 \end{figure*}

We juxtapose our estimated soiling ratio with the ground-truth soiling derate and the performance metric used in SRR. We have three different ways of estimating the soiling ratio: our  baseline approach, \mone and \mtwo, which are \st{thoroughly} described in \Cref{section:soiling}.
 In our estimates, we map negative values and values greater than one to zero and one, respectively. Then, we apply a rolling median with windows of one day.

For computing the performance metric as in SRR, we rely again on the publicly available RdTools package~\cite{rdtools}. We use as input aggregate daily values calculated on measurements taken between 12:00 and 14:00, with  irradiance greater than $500 W/m^2$. We first compute the performance metric as the ratio of   realized to modelled PV energy yield, where modelled PV energy yield is derived from a standard formula which is implemented in pvlib package~\cite{pvlib}. Then, we perform a few processing steps as suggested in RdTools' tutorials\footnote{\url{https://rdtools.readthedocs.io/en/stable/examples/degradation_and_soiling_example_pvdaq_4.html}}. We first normalize the time series with the expected power, we then apply default filters to remove clipping effects and outliers, and finally, we resample to one-day values. 

  Let $\mathcal{SD}$ be the soiling derate time series.  We denote by $\mathcal{PM}$ the performance metric used in SRR. 
 In \Cref{fig:soilingratio}, we plot our estimated soiling ratio, for all three models discussed in \Cref{section:soiling}, the soiling derate and the performance metric used in SRR, for the site of Eugene.  Compared to the other datasets, Eugene has periods of declining performance which are more apparent.  PV modules at the
Eugene site were cleaned on March 11, July 10, 
August 14, August 21, 
and August 26. 
No significant precipitation is observed during July and August, which leads to a rapid drop in the performance.
 
 We also calculate the root-mean-square error (RMSE) comparing the soiling derate with each modelled ratio, for all three sites. Since no manual cleanings were performed in Golden, the baseline algorithm and \mtwo cannot be executed. 
 We list these results in \Cref{table:rmse1}. It becomes evident, both from the RMSE values and from the visual inspection of the figure, that a better estimation of the soiling ratio can be derived by our models, compared to the model based on an analytical formula which is employed by SRR, in a setting where a soiling tendency needs to be detected, nearly real-time, on newly incoming data. Further,  \mtwo compares favourably to \mone, and improves upon the baseline algorithm in the Eugene dataset. On the other hand, both the baseline algorithm and \mtwo cannot be executed in the Golden dataset, due to the lack of manual cleanings. 
 \mone and \mtwo  present slightly diverse behaviors, rendering each potentially preferable in diverse real-world settings, depending on the exact objective of a solar park operator. Specifically, \mtwo provides the most accurate method in approximating soiling ratio, thus preferable when small to medium soiling events are tolerable by the operator, as long as ``false alarms'' are minimised. On the other hand, \mone, while slightly missing in accuracy, it is more sensitive in the detection of smaller (potential) soiling events, making it ideal in cases when even small soiling events need to be handled. Finally, we can see that the formula used in SRR essentially predicts the majority of the considered  period as soiling; a behavior with small practical use in a real-world deployment scenario.

\todo{table has changed}
\begin{table}[h]
\centering
 \caption{Evaluation.} 
 \label{table:rmse1}
 \begin{tabular}{|c|c|c|c|}
\hline
\textbf{Model} &  \multicolumn{3}{|c|}{\textbf{RMSE against $\mathcal{SD}$} } \\ 
  \cline{2-4}
 & \textbf{Eugene} & \textbf{Cocoa} & \textbf{Golden} \\ 
 \hline Baseline  & 0.006 &  0.006 & -   \\\hline
 \mone $(w_1=2,w_2=1,w_3=2)$   & 0.010 & 0.006 & 0.008 \\ \hline 
  \mone $(w_1=10,w_2=5,w_3=10)$   & 0.007 & 0.008 & 0.008 \\ \hline
  \mone $(w_1=30,w_2=10,w_3=30)$   & 0.009 & 0.007 & 0.008 \\ \hline
\mtwo  $(w_1=1,w_2=2)$& 0.008 &0.006 & - \\ 
\hline
\mtwo  $(w_1=5,w_2=10)$& 0.005 &0.007 & - \\ 
\hline
\mtwo  $(w_1=10,w_2=30)$& 0.007 &0.007 & - \\ 
\hline
$\mathcal{PM}$ used in SRR & 0.019 & 0.020 & 0.028\\ \hline
 \end{tabular}

\end{table}

 \begin{figure}[h]
     \centering
     \includegraphics[width=\columnwidth]{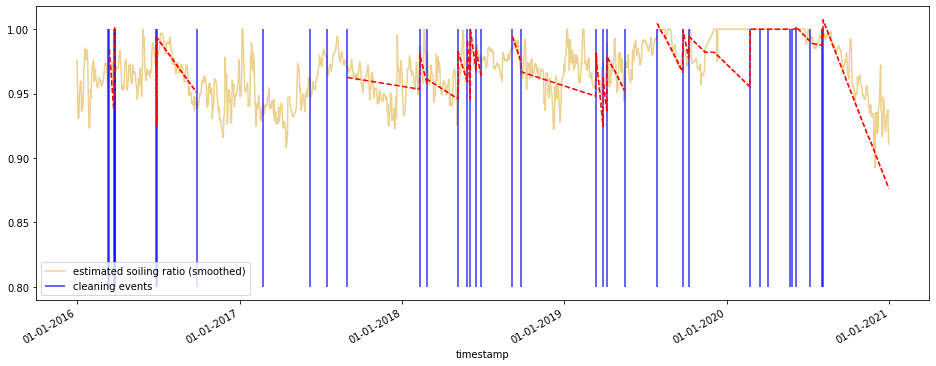}
     \caption{{Segmentation and estimated soiling ratio obtained by \mone.} \st{The yellow curve corresponds to our estimated soiling ratio obtained by training our regression model in periods following the changepoints. For illustrative purposes, a rolling median of $4$ days has been applied.
     Red dotted lines correspond to the lines obtained by the Theil-Sen method on the estimated soiling ratio.} }
     \label{fig:inaccess1}
 \end{figure}

  \begin{figure}[h]
     \centering
     \includegraphics[width=\columnwidth]{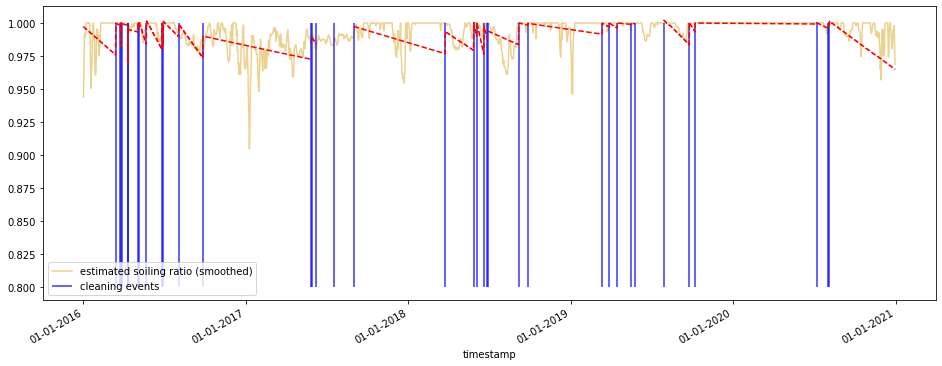}
     \caption{{Segmentation and estimated soiling ratio obtained by \mtwo.} \st{The yellow curve corresponds to our estimated soiling ratio obtained by training our regression model in periods following the changepoints. For illustrative purposes, a rolling median of $4$ days has been applied. 
     Red dotted lines correspond to the lines obtained by the Theil-Sen method on the estimated soiling ratio.}}
     \label{fig:inaccess2}
 \end{figure}

\subsubsection{ Required accuracy of regression models}\label{ss:mapethreshold}
{We experimentally justify our choice of $5\%$ as a threshold for validation MAPE of our models in methods \mone $(w_1=10,w_2=5,w_3=10)$, \mtwo $(w_1=5,w_2=10)$, as discussed in Section~\ref{section:soiling}. To be able to execute both methods for various thresholds, we employ them on the two datasets that are accompanied by manual cleaning information, i.e., Eugene and Cocoa. For both methods, we calculate the mean (over the two sites) RMSE against $\mathcal{SD}$. Experiments in Table~\ref{table:modeleval} indicate that the best result is obtained for $5\%$ (or above), for \mtwo. 

 \begin{table}[h]
\centering
 \caption{Choice of validation MAPE threshold.} 
 \label{table:modeleval}
 \begin{tabular}{|c|c|c|} 
 \hline
 \textbf{MAPE threshold} & \multicolumn{2}{|c|}{\textbf{mean RMSE against $\mathcal{SD}$}} \\  \cline{2-3} 
 &\mone  & \mtwo   \\ \hline 
$3\%$ & $0.007$ & $0.009$\\\hline
 $4\%$ & $0.008$ &    $0.007$\\ \hline
$5\%$, $10\%,15\%,20\%$ & $0.008$   &  $0.006$\\ \hline
 \end{tabular}

\end{table}
}

\subsubsection{Industrial use-case (absence of ground-truth)}
\label{sss:real}
In this section, we test our methods on the dataset described in Section~\ref{sss:inaccessdata}. 
First, we apply \mone for the detection of cleaning events. 
We filter out rains with maximum precipitation of at most $0.1$ to remove noise. 
\Cref{fig:inaccess1} {(resp.~Figure~{\ref{fig:inaccess2}})} illustrates the \st{changepoints} cleaning events detected by \mone {(resp.~\mtwo)} and our modelled soiling ratio. \st{based on training periods following the changepoints \st{detected by \mone}.}   Within each interval defined by two consecutive changepoints, 
we compute a line using the Theil–Sen method~\cite{Theil1992,Sen1968} on the estimated soiling ratio (on a 15min granularity). The Theil-Sen method is a way of fitting a line to a set of points, which is robust to outliers. The line is chosen by selecting the median slope over all lines defined by pairs of points. We plot the lines with negative slope as red dotted line segments lying in the corresponding intervals, over the course of $5$ years. We also plot a smoothed version of our estimated soiling ratio, where we 
have applied a rolling median of $5$ days.

\todo{removed paragraph (repetitive info)}

In both figures, in almost all time periods defined by two consecutive changepoints, we observe that 
there is a decreasing trend in the time series for the detected period, as dictated by the slope of the line fitted by the Theil-Sen regression (red-dotted line segments). This decreasing trend ends with rain or manual cleaning, illustrated by a blue vertical line, which is detected by our method as a cleaning event. This example is an  \emph{indication} of the effectiveness and generalizability of the proposed method. Despite the lack of labels to be able to explicitly verify the result, the trend identified is consistent with soiling and it is verifiable through the effect of  washing.

\section{{Conclusion}}
{We have described a method for estimating the soiling ratio, which uses a set of easily accessed measurements from  sensors that are commonly deployed in PV parks. Our method is data-driven, in the sense that it models the optimal performance by efficiently learning it from the data, without relying on generic formulas that fail to capture the peculiarities of the site.

Estimating the soiling ratio is  useful for PV park administrators since it allows them to schedule cleaning procedures more effectively by taking into account the rate of soil accumulation and the effectiveness of past cleaning efforts without the need for frequent visual inspections or installing specialized equipment which induces extra cost and maintenance efforts. 

Our method effectively estimates the soiling ratio in historical data. Future possible directions include extending our method to forecasting soiling losses in the future, which would assist in deciding cleaning actions at a short notice.}

\section{Acknowledgements}
The authors were partially supported by the EU's Horizon 2020 Research and Innovation programme, under the grant agreement No. 957345: “MORE”. 
\bibliographystyle{alpha}
\bibliography{sample-base}



\end{document}